\pdfoutput=1
\documentclass[10pt,twocolumn,letterpaper]{article}

\usepackage[T1]{fontenc}
\usepackage[utf8]{inputenc}
\usepackage{mathptmx}
\usepackage[letterpaper,top=0.75in,bottom=0.75in,left=0.75in,right=0.75in,columnsep=0.25in]{geometry}
\usepackage{amsmath,amssymb,bm}
\usepackage{graphicx}
\usepackage{subcaption}
\usepackage[table]{xcolor}
\usepackage{booktabs,tabularx}
\usepackage{algorithm}
\usepackage{algorithmic}
\usepackage{enumitem}
\usepackage{xspace}
\usepackage{microtype}
\usepackage{fixltx2e}
\usepackage{url}
\usepackage[numbers,sort&compress]{natbib}
\usepackage{tikz}
\usetikzlibrary{arrows,positioning,fit,calc,backgrounds,decorations.pathreplacing,patterns}
\usepackage[hidelinks]{hyperref}
\hypersetup{pdftitle={ISER-MoE: Dynamic In-Switch Replication of Hot Experts for Load Balancing in Expert-Parallel},
  pdfauthor={Jintao Peng}}
\expandafter\def\expandafter\UrlBreaks\expandafter{\UrlBreaks\do\a\do\b\do\c\do\d\do\e\do\f\do\g\do\h\do\i\do\j\do\k\do\l\do\m\do\n\do\o\do\p\do\q\do\r\do\s\do\t\do\u\do\v\do\w\do\x\do\y\do\z\do\A\do\B\do\C\do\D\do\E\do\F\do\G\do\H\do\I\do\J\do\K\do\L\do\M\do\N\do\O\do\P\do\Q\do\R\do\S\do\T\do\U\do\V\do\W\do\X\do\Y\do\Z\do0\do1\do2\do3\do4\do5\do6\do7\do8\do9\do-\do_}
\setlist{leftmargin=*,itemsep=1pt,topsep=2pt,parsep=0pt}
\makeatletter
\renewcommand\section{\@startsection{section}{1}{\z@}{-2.75ex \@plus -1ex \@minus -.2ex}{1.75ex \@plus .2ex}%
  {\normalfont\Large\bfseries}}
\renewcommand\subsection{\@startsection{subsection}{2}{\z@}{-2.5ex \@plus -1ex \@minus -.2ex}{1ex \@plus .2ex}%
  {\normalfont\large\bfseries}}
\makeatother
\newif\ifdraftnotes
\draftnotesfalse

\newcommand{\sys}{ISER-MoE\xspace}
\newcommand{\sami}{SAMI\xspace}
\newcommand{\megamoe}{MegaMoE\xspace}

\newcommand{\us}{\,$\mu$s\xspace}
\newcommand{\GBps}{\,GB/s\xspace}
\newcommand{\mOneTL}{b}
\renewcommand{\mOneTL}{c} %
\newcommand{\eg}{e.g.,\xspace}

\graphicspath{{figures/}}

\title{\Large\bfseries ISER-MoE: Dynamic In-Switch Replication of Hot Experts for Load Balancing in Expert-Parallel}
\author{Jintao Peng \quad Ethan Zhang \quad Kefeng Duan \quad Bangyu Sheng\\
Min Li \quad Itamar Rosenfeld Rauch \quad Jack Yang \quad June Yang\\
NVIDIA\\
\texttt{\{jintaop, etzhang, kduan, bangyus, lmin, irauch, jackyang, juney\}@nvidia.com}}
\date{}

\begin{document}
\maketitle

\begin{abstract}
Expert-parallel mixture-of-experts (MoE) layers run at the pace of the most loaded GPU, and a few hot experts carry most of the excess load.
Replicating them at every MoE layer balances the load only if their weights arrive within the layer, and broadcasting from the owner is the bottleneck: point-to-point copies overload it, and baseline broadcasts pay latency or bandwidth overheads.
We present \sys (In-Switch Expert Replication for MoE), built in TensorRT-LLM, which multicasts hot experts through the NVLink switch to fixed GPU groups, so each byte leaves its owner once.
A GPU scheduler assigns hot experts to groups per layer, and a multi-port broadcast spreads the injection over the group's ports, both overlapped with the shared expert.
In trace replay, the scheduler removed 98--99\% of the excess load; in prefill serving of DeepSeek-V4-Pro, throughput rose 17.4\% on a math benchmark and 7.2\% on agentic prompts.
\looseness=-1\par\end{abstract}

\section{Introduction}\label{sec:intro}

Agentic workloads send long, multi-turn prompts to large mixture-of-experts (MoE) models~\cite{aa-agentperf,sa-agentx}, and under disaggregated serving~\cite{distserve,splitwise} their prefill holds most of the GPUs: published agentic configurations of DeepSeek-V4-Pro~\cite{deepseekv4} on GB300 use 24 context (prefill) GPUs against 8 or 16 for generation~\cite{trtllm-blog-dsv4}.
Expert parallelism (EP) spreads each MoE layer's routed experts over $P$ GPUs (ranks; EP8 denotes $P=8$), and fused MoE kernels such as \megamoe~\cite{deepgemm,trtllm-megamoe} synchronize all ranks, so each layer runs at the pace of its most loaded rank.
On agentic coding traces, the rank imbalance ratio $R$ of DeepSeek-V4-Pro prefill, the most loaded rank's routed tokens over the mean, averaged 1.21--1.45 at EP4--8, and 1.69 at EP8 on GSM8K, the Grade School Math 8K benchmark (Section~\ref{sec:bg-imbalance}).
Training and reinforcement-learning post-training face the same imbalance~\cite{ultraep,moonep,foremoe}; the method applies in principle to prefill, generation and training, and we implement and evaluate it for prefill.

A few hot experts carry most of each layer's excess (Section~\ref{sec:obs-longtail}), so replicas in spare \emph{helper slots} on underloaded ranks can absorb it.
Static history-based plans~\cite{deepseekv3,deepseek-eplb} remove standing skew but not its drift between batches, so recent systems replicate hot experts at every MoE layer invocation: UltraEP~\cite{ultraep} and MoonEP~\cite{moonep} plan replicas on the GPU from the current routing, and the concurrent MegaFlux~\cite{megaflux} does so inside the \megamoe kernel that we also extend.

Replication must move whole experts (37.16\,MB each for DeepSeek-V4-Pro in NVFP4, NVIDIA's 4-bit floating-point (FP4) format) within the layer, which raises three problems.
\emph{Motivation~1} (Section~\ref{sec:mot-budget}): in a peak-rate model of FP4 prefill on GB300 (EP8), routed-expert computation hides only one or two DeepSeek-V4-Pro expert copies, whereas the shared expert, which needs no token traffic, leaves room for copies to all four helper slots%
{}, and on VR200 (Vera Rubin NVL72) for copies in 11 of 16 modeled scenarios instead of none%
.
\emph{Motivation~2} (Section~\ref{sec:mot-owner}): a hot expert needs several replicas from one owner, so point-to-point fan-out makes the owner a hot spot.
\emph{Motivation~3} (Section~\ref{sec:mot-structures}): baseline broadcasts pay latency or bandwidth overheads, from rings, from moving the payload twice in scatter + all-gather, or from the single injecting port of in-switch multicast.

In-switch multicast, NVLink's Scalable Hierarchical Aggregation and Reduction Protocol (NVLink SHARP, NVLS), removes the owner hot spot: the switch replicates each store into every member GPU~\cite{cuda-multicast,nvswitch-hotchips22}, so each byte of a hot expert leaves its owner once, whatever the replica count.
Multicast groups, however, are fixed before use, and UltraEP set in-switch multicast aside because replica groups change across layers and micro-batches~\cite{ultraep}.
\sys (In-Switch Expert Replication for MoE) fixes the groups instead of the replicas (Figure~\ref{fig:proposal}).
At each MoE layer invocation, its scheduler, \mbox{HALO-Q} (Hot-Aligned, Locality-Optimal expert replication with Quota repair), places each hot expert on one of a hierarchy of aligned groups created at initialization and reroutes part of its tokens to the replicas.
A replication kernel then multicasts the expert from its owner or through \sami (Scalable Asynchronous Multi-port In-switch broadcast), which spreads the injection over the members' ports.
Both run beside the shared expert, so the MoE kernel waits only for the plan.

In trace replay, \mbox{HALO-Q} removed 98--99\% of the excess load at EP4--24 (8,192 tokens/rank).
On GB300 at EP8, context-only (prefill-only) throughput rose 17.4\% on GSM8K and 7.2\% on agentic prompts, more than with perfectly balanced routing (13.5\% and 6.0\%).
In disaggregated serving on a GB300 NVL72 rack, it improved P90 request latency by 3.1--4.3\% and output throughput by 1.4--1.9\%.
On pre-release Rubin GPUs at EP8, context-only throughput rose 6.1\% on agentic prompts (BF16 combine). %

Our contributions are:
\begin{itemize}
  \item \textbf{Multicast weight replication} (Sections~\ref{sec:design-halom} and~\ref{sec:design-hierarchy}): to our knowledge, the first to replicate experts per invocation by in-switch multicast, at the cost of a full copy and a helper slot per member of a fixed aligned group.
  \item \textbf{\sami} (Section~\ref{sec:design-sami}): the classic scatter + all-gather broadcast~\cite{thakur2005mpich,blink} applied to weight replication, fused per slice, with the all-gather multicast from every member's port.
  \item \textbf{A shared-expert copy window} (Sections~\ref{sec:mot-budget} and~\ref{sec:design-overlap}): a peak-rate model of FP4 prefill, after the DeepSeek-V4 report and the MoE inference system PROBE~\cite{deepseekv4,probe}, indicating that the shared expert gives copies the room routed experts lack, and a runtime that hides planning and replication in it.
  \item \textbf{Artifact}: \sys in TensorRT-LLM as DynamicEPLB, a dynamic expert parallelism load balancer, in an open pull request under review~\cite{trtllm-pr19646} (Section~\ref{sec:impl}).
\end{itemize}

\section{Background and Observation}\label{sec:background}

\subsection{Expert-Parallel MoE}\label{sec:bg-ep}

\begin{figure}[t]
\centering
\includegraphics{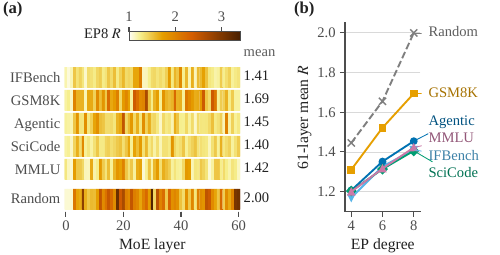}
\caption{Rank imbalance ratio $R$ (max/mean routed tokens per rank) of DeepSeek-V4-Pro prefill varies by workload, layer and EP degree (whole-trace loads, contiguous experts per rank).
(a)~EP8; right: 61-layer means. (b)~61-layer means.
Workloads: GSM8K, MMLU (Massive Multitask Language Understanding), IFBench, SciCode, agentic coding traces and random tokens.}
\label{fig:bg-imbalance}
\end{figure}

An MoE layer routes each token to $k$ of its $E$ experts~\cite{shazeer2017,gshard}.
EP spreads the experts over $P$ ranks, each holding $H=E/P$ \emph{home} experts; every invocation, in prefill, generation or training alike, \emph{dispatches} tokens to their experts' ranks and \emph{combines} the outputs back.
We target DEP, attention data parallelism (DP) with EP, in which every rank contributes its own tokens (\emph{tokens/rank}, counted before top-$k$ expansion).
Many MoE models add a \emph{shared expert} that processes every token~\cite{deepseekv4}.
Fused MoE kernels such as \megamoe~\cite{deepgemm,trtllm-megamoe} run dispatch, both expert general matrix multiplications (GEMMs) and combine in one persistent kernel that synchronizes all ranks, so every rank finishes with the most loaded one.

Per layer, the \emph{load} $L_r$ of rank $r$ counts the token--expert assignments (\emph{routes}) it processes; the \emph{rank imbalance ratio} $R=\max_r L_r/\bar{L}$ lies between 1 and $P$, $R-1$ is the \emph{excess}, and ranks above (below) $\bar{L}$ are \emph{hot} (\emph{cold}).
A \emph{replica} copies a hot (heavily routed) expert into a \emph{helper slot}, a per-rank buffer for one more expert, and takes routes from the expert's \emph{owner}, its home rank.
\emph{Per-invocation} planning~\cite{ultraep,moonep,megaflux} places replicas for each layer invocation from its current batch.

\emph{In-switch multicast} (NVLS)~\cite{nvswitch-hotchips22} replicates stores inside the NVLink switches: a store to a CUDA multicast object reaches the memory of every member~\cite{cuda-multicast}.
Membership is fixed before use and only members map the object, so a writer is a member and receives its own replica; each object takes one of a limited number of switch multicast IDs (MLIDs; 1,024 per GB200 NVLink domain~\cite{gb200-partition}).
Streaming multiprocessors (SMs) issue copies through the Tensor Memory Accelerator (TMA) without the host; copy engines (CEs) need no SMs but are host-driven.

\subsection[Rank Imbalance Varies by Workload and Layer]{Rank Imbalance Varies\protect\\ by Workload and Layer}\label{sec:bg-imbalance}

\begin{figure}[t]
\centering
\includegraphics{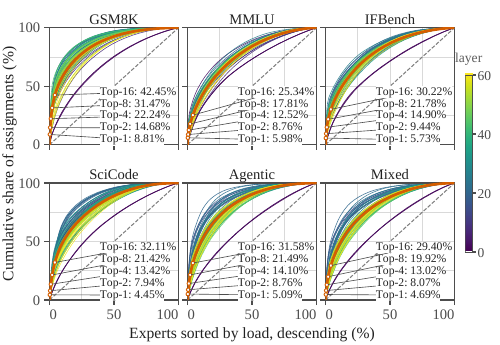}
\caption{Expert load is long-tailed (DeepSeek-V4-Pro, whole datasets): cumulative share of a layer's assignments over its 384 experts sorted by load; thin: per MoE layer; thick: 61-layer average with the shares of the Top-1 to Top-16 experts; dashed: uniform load.}
\label{fig:obs-longtail}
\end{figure}

We recorded the prefill routing of DeepSeek-V4-Pro~\cite{deepseekv4} (61 MoE layers, $E=384$, $k=6$), before load balancing, on five public workloads (GSM8K~\cite{gsm8k}, MMLU~\cite{mmlu}, IFBench~\cite{ifbench}, SciCode~\cite{scicode} and agentic coding traces from SemiAnalysis~\cite{sa-cc-traces}) and random tokens, and computed each workload's \emph{whole-trace} $R$ per layer, rank $r$ holding experts $[rH,(r+1)H)$.
$R$ depends on the workload: its 61-layer mean at EP8 was 1.40--1.45 for four workloads, 1.69 for GSM8K and 2.00 for random tokens (Figure~\ref{fig:bg-imbalance}(a)).
It depends on the layer (1.13--2.51 across the agentic traces' layers at EP8), and any two workloads share at most one of their five most imbalanced layers.
It grows with the EP degree, from 1.17--1.31 at EP4 to 1.40--1.69 at EP8 (Figure~\ref{fig:bg-imbalance}(b)).
An invocation, however, sees only its batch: in a mixed trace interleaving the five workloads by request (4,340 requests, 18.2M prefill tokens, 91\% agentic), $R$ per iteration (forward pass) of 8,192 tokens/rank averaged 1.20, 1.44, 1.84 and 2.24 at EP4, EP8, EP16 and EP24, against 1.41 whole-trace at EP8.
Most of a fixed mix's skew is thus standing, and a static plan from load history, as in DeepSeek's Expert Parallelism Load Balancer (EPLB)~\cite{deepseek-eplb}, removes most of it but not its drift between batches (Section~\ref{sec:eval-placement}).
Imbalance thus costs time: on GB200 (EP8, 8,192 tokens/rank, trace-derived routing), \megamoe without replicas took $513+1051\,R$\us over the 61 layers (Pearson correlation 0.938).

\subsection[Observation: Long-Tail Distribution Enables Hot-Expert Replication]{\protect\raggedright Observation: Long-Tail Distribution Enables Hot-Expert Replication}\label{sec:obs-longtail}

Expert load is long-tailed (Figure~\ref{fig:obs-longtail}): over the mixed trace, the four busiest of 384 experts received 13.02\% of a layer's assignments on average (1.04\% if uniform), and per iteration (EP8, 16,384 tokens/rank) the four busiest still received 13.6\%.
A hot rank's excess thus sits on a few home experts: at EP8, the four busiest home experts of a layer's hottest rank carried at least that rank's excess in 59 of the 61 layers.
Replicating them onto cold ranks can absorb most of the imbalance; \mbox{HALO-Q} therefore considers each rank's $\min(S,H)$ busiest home experts (Section~\ref{sec:design-halom}), with $S=4$ helper slots per rank by default.
Each replica, however, is a whole expert, copied within the invocation that uses it.

\section{Motivation}\label{sec:motivation}

\subsection[Motivation 1: More Weight-Copy Overhead Than Routed-Expert Computation]{\protect\raggedright Motivation 1: More Weight-Copy\protect\\ Overhead Than Routed-Expert Computation}\label{sec:mot-budget}

\begin{figure}[t]
\centering
\includegraphics{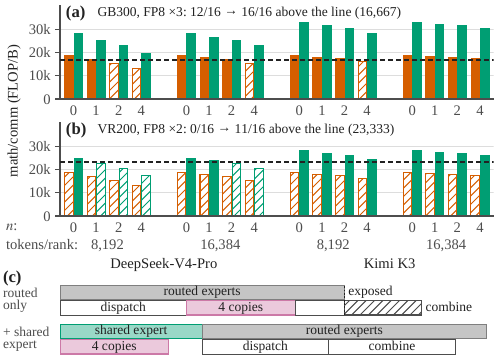}
\caption{Motivation~1: with the shared expert, the copies overlap in all 16 modeled scenarios on GB300, against 12 with routed experts only (peak-rate model per MoE layer and rank; EP8, uniform routing, NVFP4, token traffic not deduplicated, as in \megamoe).
(a)~GB300~\cite{gb300-nvl72}: math/comm ratio (FLOPs per communicated byte) with $n$ copies per rank; red: routed experts only; green: with the 8-bit floating-point (FP8) shared expert at the FP4:FP8 rate ($\times 3$); hatched: below the device ratio (dashed).
(b)~VR200~\cite{nvidia-dgx-vr-nvl72}, the same with $\times 2$. %
(\mOneTL)~Example from (a) in time (DeepSeek-V4-Pro, 8,192 tokens/rank, four copies): beside the routed experts, the copies delay the combine; beside the shared expert, they hide.
}
\label{fig:m1-ratio}
\end{figure}

Per-invocation replica copies~\cite{ultraep,moonep,megaflux} share the links with the layer's token traffic; a peak-rate model of FP4 prefill shows how many of them computation can hide.

\textbf{Math/comm ratio.}
Per MoE layer and rank, $\tau$ tokens make $\tau k$ token--expert pairs.
Each costs $6H_eI$ FLOPs in its routed expert ($H_e$ the expert's input width) and, under uniform routing, leaves the rank with probability $\frac{P-1}{P}$, moving $H_eb$ bytes in both dispatch and combine ($b=0.5625$\,B per NVFP4 element).
Each of the $n$ expert copies a rank receives or sends moves $W=3H_eIb$ bytes.
Per link direction,
\begin{equation*}
\frac{\text{math}}{\text{comm}}=\frac{6\,\tau k H_eI}{\underbrace{2\,\tfrac{P-1}{P}\,\tau k H_eb}_{\text{dispatch + combine}}+\underbrace{\,nW\,}_{\text{copies}}}.
\end{equation*}
Communication hides behind computation at peak rates only if this ratio exceeds the device ratio $F_4/B_{\mathrm{link}}$ of dense FP4 FLOP/s to per-direction link bandwidth: 16,667\,FLOP/B on GB300 (15\,PFLOP/s, 900\GBps~\cite{gb300-nvl72}).
On VR200 it is 23,333\,FLOP/B (35\,PFLOP/s and 1.5\,TB/s per GPU and direction, from the rack datasheet~\cite{nvidia-dgx-vr-nvl72}). %

\textbf{Routed experts only.}
We model DeepSeek-V4-Pro ($H_e=7168$, $k=6$) and Kimi~K3~\cite{kimi-k3}, a latent MoE (top-16 of 896 experts, $H_e=3584$); both have $I=3072$ and are assumed NVFP4 (Kimi~K3 exceeds our implementation's 384-expert limit and is only modeled).
Without copies the ratio is $3I/b\cdot P/(P-1)=18{,}725$\,FLOP/B at EP8, whatever $\tau$, $H_e$ and $k$; each copy lowers it, and of 16 scenarios (0, 1, 2 or 4 copies; 8,192 or 16,384 tokens/rank), 12 stay above GB300's ratio (Figure~\ref{fig:m1-ratio}(a), red).
At 8,192 tokens/rank, DeepSeek-V4-Pro's routed computation (432.9\us) exceeds its dispatch plus combine (385.4\us) by only 1.15 copies of 41.3\us; four copies would outlast the computation by 117.6\us and delay the combine (Figure~\ref{fig:m1-ratio}(\mOneTL)); at 16,384 tokens/rank, two hide.

\textbf{Counting the shared expert.}
The shared expert needs neither the replica plan nor token communication, so it can run before dispatch, while the links are idle.
Weighting its FP8 FLOPs by $F_4/F_8=3$ on GB300 (assuming FP8 shared experts in both models), the ratio rises to 19,661--32,768\,FLOP/B and all 16 scenarios allow overlap, four copies per rank included (Figure~\ref{fig:m1-ratio}(a), green).
On VR200, where FP8 runs at half the FP4 rate, the shared expert raises the count from 0 to 11 (Figure~\ref{fig:m1-ratio}(b)), but DeepSeek-V4-Pro still hides at most one copy. %
On GB300, DeepSeek-V4-Pro's shared expert lasts 216.5\us at 8,192 tokens/rank, room for about five more copies at line rate.
Copies should therefore start right after routing, beside the shared expert, not at the entry of \megamoe (Figure~\ref{fig:proposal}(a)).

\begin{figure}[t]
\centering
\input{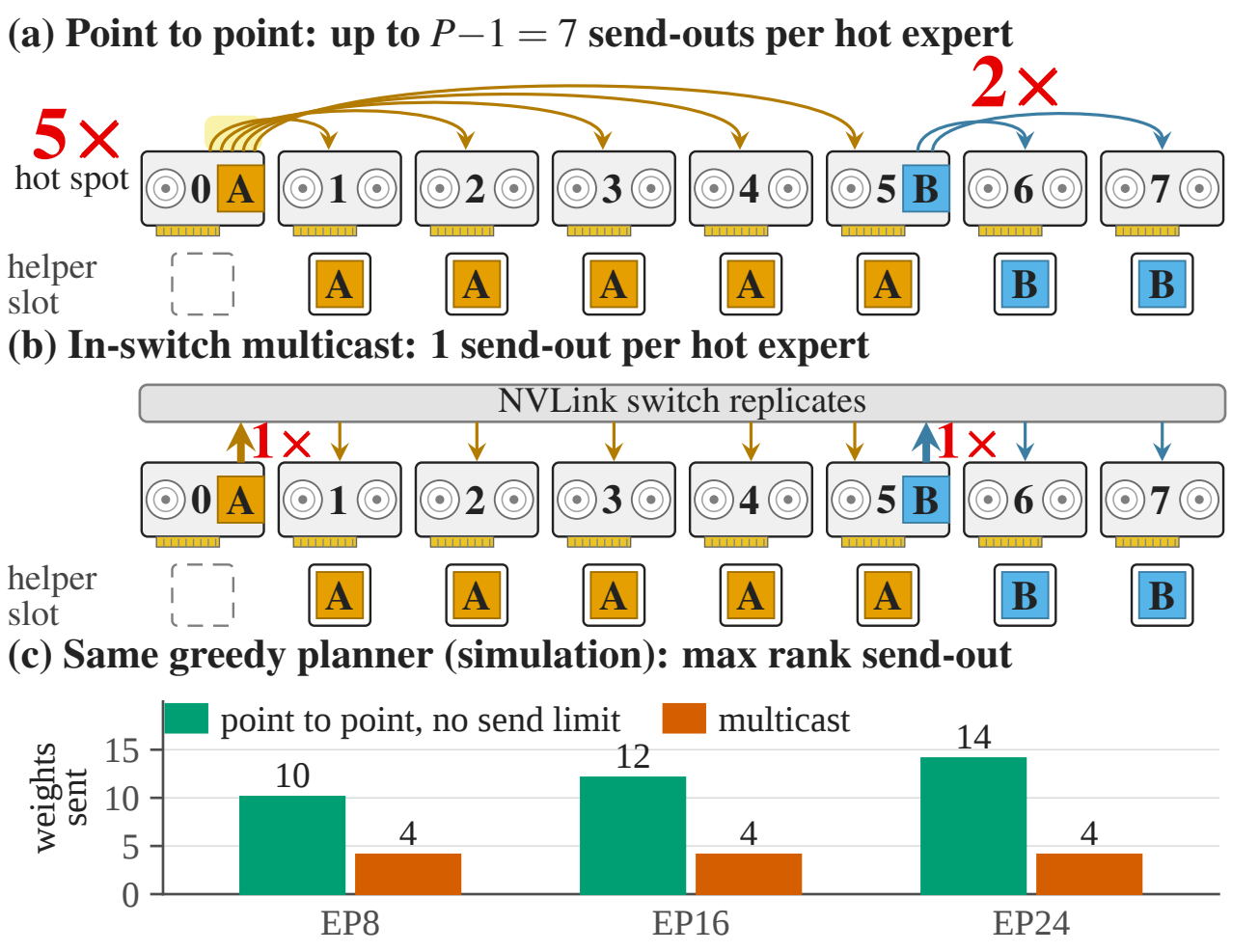}
\caption[Source broadcast hot spot]{Replication makes the owner a broadcast hot spot. (a,\,b)~EP8, idealized: hot experts A on GPU~0 and B on GPU~5 need five and two replicas, one per helper slot (below each GPU). (c)~Simulated trace replay of a greedy planner (DeepSeek-V4-Pro, 8,192 tokens/rank, $S{=}4$), point to point or multicast: the busiest rank's worst-case send-out per layer invocation.}
\label{fig:m2-hotspot}
\end{figure}

\begin{figure*}[t]
\centering
\input{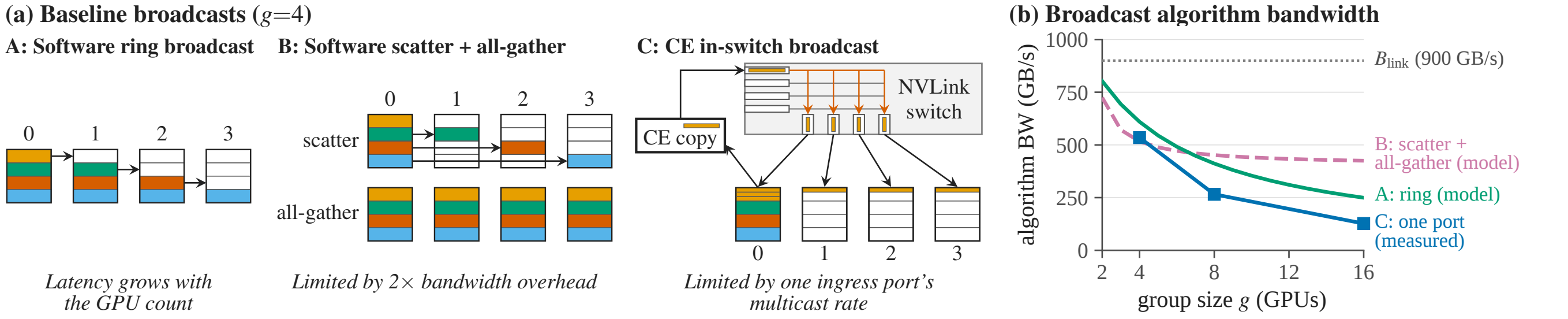}
\caption[Baseline broadcasts]{Baseline broadcasts are limited by latency or bandwidth. (a)~One expert's weights, in four slices, from GPU~0 to $g{=}4$ GPUs. (b)~Algorithm bandwidth (BW; payload/time) of broadcasting to $g$ GPUs: A and B modeled for one expert (37.16\,MB) with $B_{\mathrm{link}}=900$\GBps, $\alpha=5\,\mu$s and 2-MiB ring slices; C measured on four GB200 trays (160\,MiB).}
\label{fig:m3-bcast}
\end{figure*}

\textbf{Limits of the model.}
The model assumes peak rates, uniform routing and an NVFP4 combine; with the implementation's default 16-bit brain floating-point (BF16) combine, token traffic alone outlasts DeepSeek-V4-Pro's routed experts on GB300 (685 vs.\ 433\us at 8,192 tokens/rank), so copies hide only beside the shared expert.
It also counts helper tiles, which cannot hide their own copy (Section~\ref{sec:design-overlap}), as computation.
At the 166--450\GBps of the replication kernel on four GB200 SMs (Section~\ref{sec:eval-bcast}), four copies take 330--898\us and spill into the routed phase, which R1 allows (Section~\ref{sec:mot-req}).

\subsection{Motivation 2: Source Broadcast Hot Spot}\label{sec:mot-owner}

Hot experts often need several replicas each, and since they are a hot rank's busiest home experts (Section~\ref{sec:obs-longtail}), one owner exports several of them, while each receiver takes at most one per helper slot.
If an owner replicates $K$ experts, expert $e$ to $d_e$ ranks, point-to-point copies send $\sum_e d_e\le K(P-1)$ payloads out of the owner; relays, as in UltraEP, which identified and modeled this fan-out bottleneck~\cite{ultraep}, relieve the owner at the cost of a store-and-forward hop but still inject $\sum_e d_e$ in total.
In-switch multicast sends each byte out of the owner once, $K$ payloads, with no extra hop (Figure~\ref{fig:m2-hotspot}(a,\,b)).

We replayed the mixed trace (8,192 tokens/rank, $S=4$; simulated assignment load) through a greedy point-to-point planner modeled on UltraEP and MegaFlux~\cite{ultraep,megaflux}: like the placement stage of \mbox{HALO-Q} (Section~\ref{sec:design-halom}), it moves the hottest rank's busiest home experts to the coldest ranks, but copies each replica directly from its owner to any below-mean rank with a free slot.
Unlimited, it balanced slightly better than \mbox{HALO-Q} ($R=1.001$--1.009 against 1.006--1.020 at EP8--24), but its busiest owner sent 7/9/10 payloads per invocation at EP8/16/24 at the 99th percentile (Figure~\ref{fig:m2-hotspot}(c)), while no rank received more than four; at line rate, these 99th-percentile send-outs take 289--413\us, longer than the 216.5\us shared-expert window.
Limiting each owner to four copies removed this tail but raised $R$ to 1.016--1.100, whereas \mbox{HALO-Q}'s multicast plans kept the busiest owner at a median of 2/3/3 payloads, never above four.

\subsection[Motivation 3: Baseline Broadcasts Are Limited by Latency and Bandwidth Overhead]{\protect\raggedright Motivation 3: Baseline Broadcasts Are\protect\\ Limited by Latency and Bandwidth Overhead}\label{sec:mot-structures}

Multicast lets each byte leave the owner once, but each of three ways to broadcast an expert to a group of $g$ GPUs pays a latency or bandwidth overhead (Figure~\ref{fig:m3-bcast}(a), A--C).
We model the algorithm bandwidth of A and B from the per-direction link bandwidth $B_{\mathrm{link}}$ and a latency $\alpha$ per step, and measure C (Figure~\ref{fig:m3-bcast}(b)).

\textbf{A: software ring.}
The NVIDIA Collective Communications Library (NCCL)~2.31.2 broadcasts only along a pipelined ring~\cite{nccl,nccl-demystify}.
Each link carries the payload once, but $c$-byte slices add a pipeline fill that grows with the group, $T_A=W/B_{\mathrm{link}}+(g-1)\alpha+(g-2)c/B_{\mathrm{link}}$, so the ring's bandwidth falls to 250\GBps at $g=16$.

\textbf{B: software scatter + all-gather.}
The classic large-message broadcast~\cite{thakur2005mpich} spreads injection over all $g$ ports but moves the payload twice in series, $T_B=2\frac{g-1}{g}\,W/B_{\mathrm{link}}+2\alpha$, which caps its bandwidth near $B_{\mathrm{link}}/2$.

\textbf{C: CE in-switch broadcast.}
A CE writer injects each byte once and the switch makes all $g$ replicas, but from one ingress port replicated at a fixed rate: on four GB200 trays its total output stayed at 2.0--2.1\,TB/s, so each receiver got 535, 266 and 127\GBps at $g=4$, 8 and 16.
One injection per byte is thus necessary but not sufficient; injection must also be spread over many ports.

\subsection{Design Requirements}\label{sec:mot-req}

\begin{figure*}[t]
\centering
\input{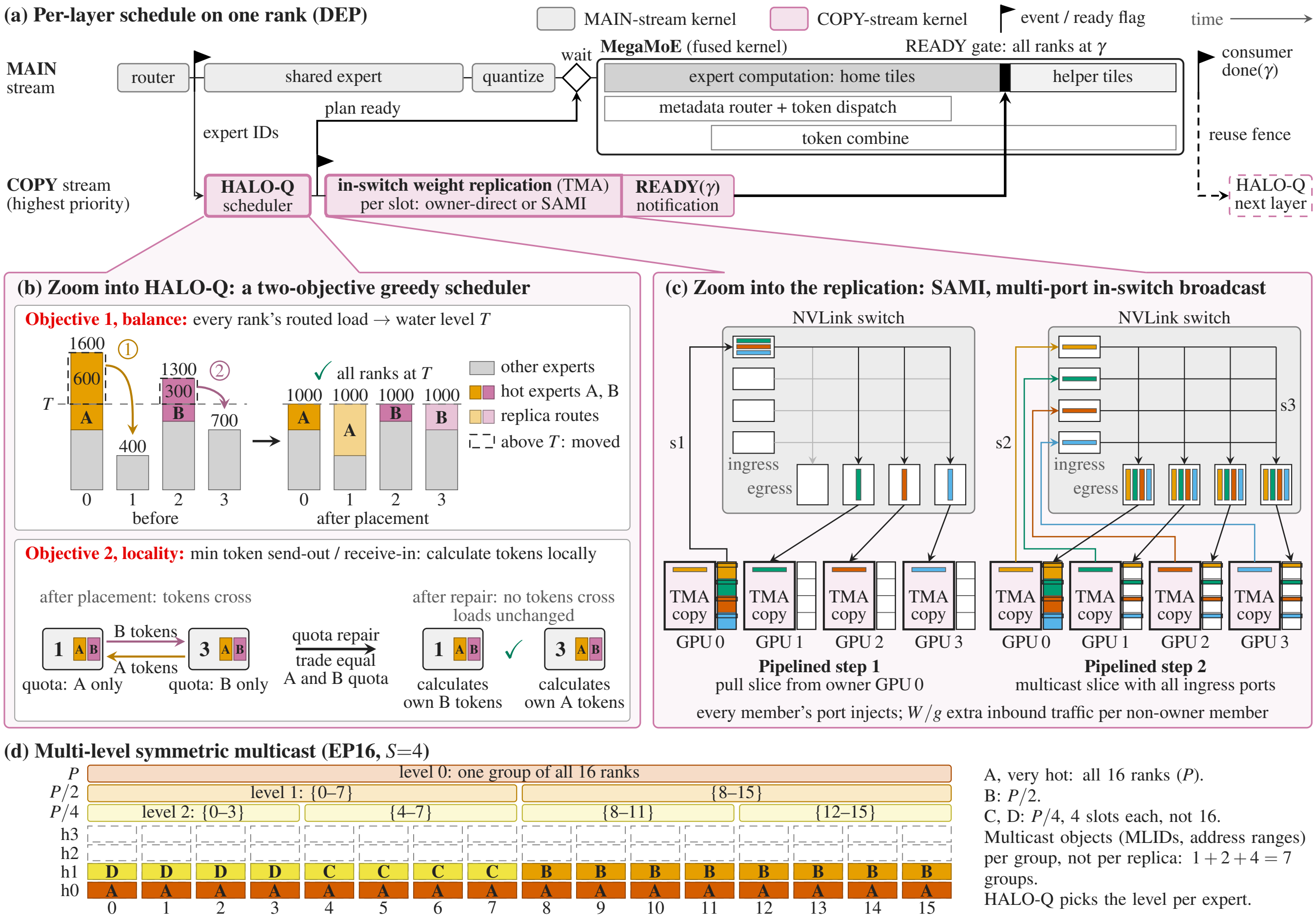}
\caption{\sys in one MoE layer (schematic).
(a)~One rank (DEP): after routing, the highest-priority COPY stream runs \mbox{HALO-Q} and the replication kernel beside the
shared expert; \megamoe waits only for the plan, and only its helper tiles wait, at the READY gate, for
generation~$\gamma$ (a per-invocation counter) from every rank.
(b)~\mbox{HALO-Q} (ranks 0--3).
Objective~1, balance: the busiest rank's hottest expert gets a replica (steps 1, 2), and the coldest ranks take its
routes above the water level~$T$, the mean rank load.
Objective~2, locality: after placement, ranks~1 and~3 would exchange tokens of the multicast experts B and A; quota repair
trades equal parts of the two quotas, so both calculate their own tokens locally, at unchanged loads.
(c)~Replication with \sami ($g=4$, owner GPU~0) in two pipelined steps: each member pulls its slice from
GPU~0 (s1) and stores it through its own ingress port (s2), and the switch multicasts it to all members (s3).
(d)~Multi-level symmetric multicast: symmetric groups of $P$, $P/2$ and $P/4$ ranks, each with multicast objects;
\mbox{HALO-Q} picks the level per expert.
Rows h0--h3: helper slots.}
\label{fig:proposal}
\end{figure*}

The three motivations yield four requirements.
\textbf{R1} (Motivation~1): start planning and copying right after routing, hide them behind the shared expert and then the home-expert computation, and make only helper work wait for replica weights.
\textbf{R2} (Motivation~2): bound the experts each owner exports per layer, since even multicast sends each out of the owner once.
\textbf{R3} (Motivation~2): each byte of a hot expert leaves its owner once, whatever the replica count; as multicast membership is fixed before use and stores land at the same offset on all members (Section~\ref{sec:bg-ep}), this requires fixed groups and symmetric helper slots.
\textbf{R4} (Motivation~3): spread injection over many ports, by distinct owners or by group members, also when the owner lies outside the group.
Sections~\ref{sec:design-overlap}, \ref{sec:design-halom}, \ref{sec:design-hierarchy} and~\ref{sec:design-sami} meet R1--R4 in order.

\section{\sys Design}\label{sec:design}

Figure~\ref{fig:proposal} shows one MoE layer under \sys: after routing, the \mbox{HALO-Q} scheduler decides which hot
experts to replicate where and maps every route to an expert slot (Section~\ref{sec:design-halom}), and a replication
kernel multicasts the replicas into aligned rank groups (Sections~\ref{sec:design-hierarchy} and~\ref{sec:design-sami}),
beside the shared expert (Section~\ref{sec:design-overlap}).

\subsection{HALO-Q Scheduler}\label{sec:design-halom}\label{sec:design-repair}

\mbox{HALO-Q} (Hot-Aligned, Locality-Optimal expert replication with Quota repair) is one GPU kernel per MoE layer
invocation.
It replicates hot experts in aligned rank groups (Section~\ref{sec:design-hierarchy}) with two objectives, in order
(Figure~\ref{fig:proposal}(b)):
\begin{itemize}
\item \textbf{Objective 1, balance.}
\megamoe finishes a layer when its most loaded rank does (Section~\ref{sec:bg-imbalance}), so \mbox{HALO-Q} brings every
rank's routed load down to the water level $T$.
\item \textbf{Objective 2, locality.}
A route served off its source rank costs a token sent in dispatch and returned in combine, and the busiest
sender or receiver bounds that traffic.
Among balanced routings, \mbox{HALO-Q} therefore calculates as many tokens as possible locally.
\end{itemize}

\begin{algorithm}[t]
\caption{\mbox{HALO-Q} on EP rank $s$, once per MoE layer invocation; all ranks compute the same plan from the same
all-gathered counts. \emph{Spare} quota: quota beyond own routes.}
\label{alg:halo}
\footnotesize
\setboolean{ALC@noend}{true}%
\renewcommand{\algorithmiccomment}[1]{\hfill$\triangleright$~#1}
\begin{algorithmic}[1]
\REQUIRE routes $\mathcal{R}_s$; $P$ ranks with $H$ home and $S$ helper slots; aligned groups
\ENSURE broadcasts $b=(e_b,G_b,h_b)$; a slot per route (home unless rerouted)
\STATE $C\gets\textsc{AllGather}$(counts of $\mathcal{R}_s$); loads $L$; $T\gets\lceil\sum_r L[r]/P\rceil$
  \COMMENT{stage 1}
\WHILE[stage 2: Objective 1]{a rank above $T$ can still export}
\STATE $o\gets$ the busiest such rank; $e\gets$ its heaviest unexported home expert
\STATE $G\gets$ the aligned group, at the finest level able to absorb $e$ (else coarser), with a free slot on
  every member and the most room below $T$; if none has room, freeze $o$
\STATE otherwise $b\gets(e,G)$ with quotas $q_b$: move up to $L[o]-T$ of $e$'s routes from $o$ to $G\setminus\{o\}$,
  coldest first, each up to $T$
\ENDWHILE
\STATE $h_b\gets$ the first index free on all of $G_b$, coarse levels first \COMMENT{stage 3}
\FOR[stage 4: Objective 2]{all rank pairs $(a,c)$, twice, in both orders}
\STATE on experts both host, $a$ trades spare quota with $c$ in equal amounts, as much as both allow: it takes what
  its own routes need and gives what they do not
\ENDFOR
\FORALL[stage 5]{broadcasts $b$}
\STATE $s$ keeps up to $q_b[s]$ of its routes to $e_b$; the surplus routes of all ranks fill the hosts' spare quotas
  in rank order
\STATE a route runs in helper slot $h_b$ of its host (the home slot if the host is the owner outside $G_b$)
\ENDFOR
\end{algorithmic}
\end{algorithm}

The five stages are (Algorithm~\ref{alg:halo}):
\begin{enumerate}
\item \textbf{Count exchange.}
Ranks exchange their $E$ route counts through a symmetric NVLink buffer, with no host step, and run the same
deterministic planner, so all obtain the same plan without broadcasting it, as in UltraEP~\cite{ultraep}.
\item \textbf{Placement (Objective 1).}
With $T=\lceil\sum_r L_r/P\rceil$, \mbox{HALO-Q} repeatedly takes the busiest rank above $T$ that can still export
and its heaviest home expert not yet exported.
An owner exports at most $S$ experts per invocation, chosen among its $\min(S,H)$ heaviest, which bounds its egress
(Section~\ref{sec:mot-owner}); a rank receives at most $S$ replicas.
The expert goes to the aligned group, not necessarily containing the owner, whose members all have a free helper slot,
at the finest level that can absorb the expert at the water level (coarser only if none there has room), with the most
room below $T$; up to the owner's excess over $T$ of its routes move there, coldest member first, each up to~$T$.
An owner for which no group has room, or which has no candidate left, is frozen.
\item \textbf{Slot coloring.}
Each placed expert gets one helper index free on all members of its group (Section~\ref{sec:design-coloring}).
\item \textbf{Quota repair (Objective 2).}
Placement fixes loads and each replicated expert's total quota, not its split among hosts, so rank $a$ may send tokens of one
expert to rank $c$ while $c$ sends tokens of another to $a$.
In two rounds of a circle-method tournament over disjoint rank pairs, each rank takes, from its partner's
\emph{spare} quota (beyond the partner's own routes), quota for its own routes of an expert both host, and hands the
partner as much of its own spare quota; run in both orders, this cancels opposite flows as far as the loads allow,
while every load and total, hence balance and copy count, stays.
\item \textbf{Route rewrite (local first).}
Without communication, each rank first fills its own quota; the surplus routes of all ranks, in rank order, then fill
the other hosts' spare quotas in rank order (a north-west-corner assignment), following UltraEP's local-first
cumulative quotas~\cite{ultraep} and MoonEP's prefix sums~\cite{moonep}.
A route runs in its host's helper slot, or in the home slot if the host is an owner outside the group.
\end{enumerate}
Replicas are byte copies and gate weights are untouched, so replication changes where a route runs, not what it computes.

\subsection{Multi-Level Symmetric Multicast}\label{sec:design-hierarchy}\label{sec:groups}

Replicating a moderately hot expert on all $P$ ranks wastes helper slots, yet arbitrary groups cannot be prepared:
multicast membership is fixed before use, MLIDs are limited (Section~\ref{sec:bg-ep}), and the ranks needing a replica
change with every layer and batch.
\sys therefore fixes the groups, not the replicas: level $\ell$ splits the $P$ ranks into $2^\ell$ aligned groups of
$g_\ell=P/2^\ell$ consecutive ranks, for $\ell=0$ and every finer level with $2^\ell \mid P$ and $g_\ell\ge4$ (\eg 24,
12 and 6 ranks at EP24), and \mbox{HALO-Q} picks a level per expert (Figure~\ref{fig:proposal}(d)).
The four-rank floor is a software choice, not a multicast limit: imbalance falls as each rank holds more experts
(Section~\ref{sec:bg-imbalance}), and smaller groups are unevaluated.
All groups' multicast objects are created at initialization.

Each rank holds its $H=E/P$ home experts in local slots $[0,H)$ and $S$ helper slots after them.
A \emph{broadcast} installs an expert's weights and block scales (its \emph{bundle}) in one helper slot on every
member of an aligned group.

\textbf{Symmetric helper slots.}\label{sec:design-coloring}
A multicast store lands at the same offset on every member, so a broadcast needs one helper index free on all members.
Alignment guarantees one: if every rank belongs to the groups of at most $S$ broadcasts, coloring broadcasts from coarse
to fine, each with the smallest index free on all its members, never fails.
Aligned groups are nested or disjoint, so every earlier, equally coarse or coarser group that meets a group $G$ contains
it, and all members of $G$ see the same at most $S-1$ used indices.
\mbox{HALO-Q} ensures the precondition by admitting a broadcast only if every member has a free slot.
Without alignment, $S{+}1$ pairwise-overlapping groups would need $S{+}1$ indices.

\subsection[TMA Weight Replication and SAMI Multi-Port Broadcast]{\protect\raggedright TMA Weight Replication and\protect\\ \sami Multi-Port Broadcast}\label{sec:design-sami}\label{sec:sami}

The replication kernel copies with TMA from SMs, avoiding the host round trip of CEs: warps stage 8-KiB slices in
shared memory and store them to the group's multicast address (\texttt{multimem.\allowbreak cp.async.bulk}), so the
switch writes each slice into every member's helper slot.
\emph{Owner-direct} replication has the owner, a group member, multicast the whole bundle through its own port.
With \emph{\sami} (Figure~\ref{fig:proposal}(c)), the $g$ members share the injection in two pipelined steps, with no
barrier or reassembly: member $i$ (s1)~pulls each slice $j\equiv i \pmod g$ from the owner and (s2)~stores it through
its own port to the multicast address, and (s3)~the switch delivers it to every member.
\sami is thus a scatter + all-gather broadcast~\cite{thakur2005mpich,blink} whose all-gather is in-switch multicast, as
in NVLS AllGather~\cite{nccl}.
For an expert of $W$ bytes, the owner still sends $W$ and each member injects $W/g$, so the output fed by one port,
which limits single-port multicast (Section~\ref{sec:mot-structures}), falls from $gW$ to $W$ and the issue work is
shared; the price is $W/g$ of extra inbound traffic per non-owner member.
Since a multicast writer must be a member, \sami also serves owners outside the group, which only serve the pulls.

A slot is replicated owner-direct if its owner belongs to the group $G$ and either $G$ is \emph{pure} (every broadcast
that meets $G$ targets exactly $G$ with its owner inside) or this is the owner's widest broadcast into a group
containing it; otherwise \sami replicates it.
This default rule avoids \sami's extra inbound traffic where the owner can serve its group, relying on distinct owners
to spread the injection; whether it pays off with many owners is untested (Section~\ref{sec:eval-bcast}).

\subsection{Runtime Overlap}\label{sec:design-overlap}

\sys issues the copy right after routing, where Section~\ref{sec:mot-budget} found room beside the shared expert
(Figure~\ref{fig:proposal}(a)):
\begin{itemize}
\item The COPY stream, the model's highest-priority stream, runs \mbox{HALO-Q} and the replication kernel while the
MAIN stream computes the shared expert and quantizes the routed input.
\item MAIN then waits on the device for the plan and launches \megamoe on the rewritten routes; dispatch never waits for
the copy.
\item \megamoe claims tiles in slot order, so dispatch overlaps the home tiles, and combine the later home and the
helper tiles.
\item Only helper tiles wait: after its writes drain, the replication kernel multicasts the invocation's READY generation
to every rank, even with nothing to copy, and each cooperative thread array (CTA) waits before its first helper tile
until every rank's generation is current.
\end{itemize}
To first order, the copy is exposed only if it outlasts the shared expert and the home tiles; a copy issued inside
\megamoe, as described for MegaFlux~\cite{megaflux}, would miss the shared-expert window.

\section{Implementation}\label{sec:impl}

\sys is implemented in TensorRT-LLM as Dynamic\-EPLB, in an open pull request under review~\cite{trtllm-pr19646}; we
describe its commit \texttt{99dac8f1}.
A model enables this per-layer dynamic mode through \texttt{moe\_config.\allowbreak load\_balancer}, with
\texttt{num\_slots}${}=P(H+S)$ and an SM budget, \texttt{auxiliary\_sms}.

\textbf{Kernels, streams and SM budgets.}\label{sec:impl-streams}
We extend TensorRT-LLM's NVFP4 \megamoe kernel~\cite{trtllm-megamoe}, written in the CuTe domain-specific language, with
helper slots and the READY gate.
Its per-expert weight and scale planes append, after the $H$ home rows, one of two helper banks shared by all layers
(layer $i$ uses bank $i \bmod 2$); as in UltraEP and MoonEP~\cite{ultraep,moonep}, sharing cuts helper storage to
297\,MB per rank at $S=4$ (computed), from 9.07\,GB for 61 per-layer slot sets.
Before each \mbox{HALO-Q} launch, COPY waits until \megamoe has consumed the previous replicas; since every plan starts
with the all-rank count exchange, no copy overwrites a bank another rank still reads.

All MoE layers share the COPY stream, whose two kernels run back to back: \mbox{HALO-Q} on \texttt{auxiliary\_sms}
CTAs (default 8) and the replication kernel on 8.
The budget does not partition the GPU: the routed-input quantizer uses the remaining SMs, and \megamoe the full grid.
The shared expert is the standard gated multilayer perceptron (MLP) with no SM budget, so its overlap with COPY rests on stream priority.
An SM-budgeted fused shared expert exists only for one other GPU generation. %
Because persistent \megamoe CTAs spin at the READY gate, the copy must obtain SMs while \megamoe runs; this rests on
scheduling (early enqueue, highest priority, 8 independent CTAs), not on a hardware guarantee, and a starved copy would
hit the gate's 120-s timeout and fail rather than hang.
Stale or failed plans, malformed copy plans, gate errors and timeouts abort the job, with no fallback to home experts.

\textbf{Scope.}\label{sec:impl-scope}
The mode requires the \megamoe NVFP4 backend, DEP, an even EP degree up to 32, $E\le384$ divisible by $P$, CUDA~13.1 or
later and one multicast-capable NVLink domain.
CUDA graphs are unsupported, because the runtime binds one MAIN submission stream, orders COPY with events and keeps
per-submission host state.
No test launches \mbox{HALO-Q} or the replication kernel, and the generation phase is untested.

\section{Evaluation}\label{sec:eval}

\begin{figure}[t]
\centering
\includegraphics[width=\columnwidth]{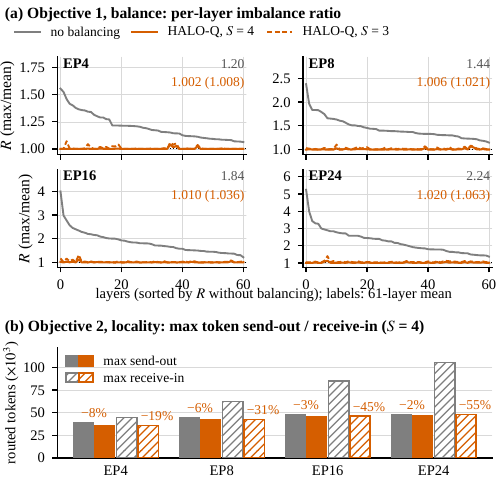}
\caption{The two objectives of \mbox{HALO-Q} in simulated trace replay (DeepSeek-V4-Pro, 8,192 tokens/rank). (a)~Objective~1, balance: per-layer $R$ without balancing, layers sorted by $R$ (dotted: $R = 1$); labels: 61-layer means without balancing and with \mbox{HALO-Q}, $S = 4$ ($S = 3$). (b)~Objective~2, locality ($S = 4$): the most routed tokens any rank sends out or receives in during dispatch, mean per iteration and layer; labels: change from no balancing.}
\label{fig:eval-placement}
\end{figure}

\begin{figure*}[t]
\centering
\includegraphics[width=\textwidth]{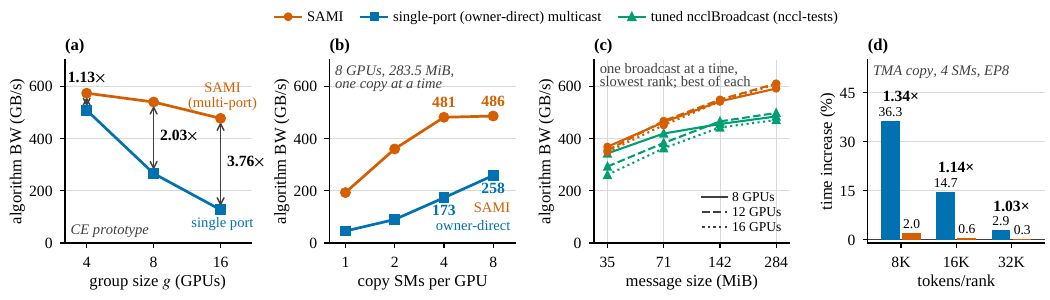}
\caption{Reference data (E2): broadcasts from one owner on GB200.
(a)~CE prototype: algorithm bandwidth of \sami (multi-port) and single-port multicast vs.\ group size $g$ ($g = 4$: two trays, 144\,MiB; $g = 8, 16$: four trays, 160\,MiB); labels: \sami's speedup.
(b)~A development build of the replication kernel (four SMs by default), standalone, slowest rank: algorithm bandwidth vs.\ copy SMs per GPU, eight 37.16\,MB bundles (283.5\,MiB, (c)'s largest size) from one owner to 8 GPUs, the owner included.
(c)~One broadcast at a time, until the slowest rank holds it: the fastest \sami variant (mostly with the owner outside the group) vs.\ the fastest tuned ncclBroadcast setting (\mbox{nccl-tests}, NCCL~2.31.2: best of protocol, buffer size, channel count, buffer registration and CUDA-graph launch); line style: GPU count.
(d)~Time increase from the shared expert's start to the end of \megamoe when the build of (b) copies four experts ($4 \times 36$\,MiB) during the shared expert (EP8); labels: \sami's speedup.}
\label{fig:eval-bcast}
\end{figure*}

\textbf{Methodology.}
All experiments use DeepSeek-V4-Pro's MoE layer with NVFP4 routed experts (Section~\ref{sec:bg-ep}).
A GB200 compute tray holds four GPUs, so EP8 spans two trays.
OFF denotes a layer without helper slots and ON the same layer with $S$ helper slots per rank.
Unless stated otherwise, kernel times are the slowest rank's median over 20 iterations, reported as the time reduction (OFF$-$ON)/OFF or the speedup OFF/ON.

\subsection{E1: Simulated Scheduler Rank Imbalance}\label{sec:eval-placement}

\begin{figure*}[t]
\centering
\includegraphics[width=\textwidth]{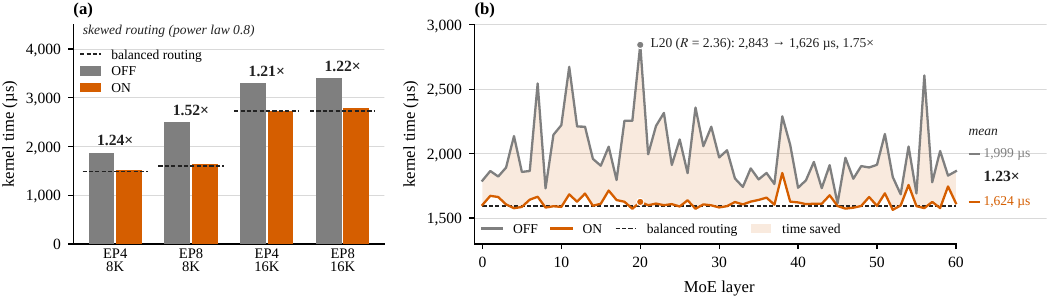}
\caption{\megamoe on GB200 (DeepSeek-V4-Pro layer, NVFP4, FP4 combine, four helper slots; slowest rank, median of 20 iterations; replicas and routes from an earlier all-rank policy, installed before timing).
(a)~Power-law routing (exponent 0.8); 8K and 16K are tokens/rank; labels: speedup.
(b)~61 layers, routes sampled from the trace (EP8, 8,192 tokens/rank); shading: time saved; margin: means; L20: layer 20.
Dashes: balanced routing (OFF).}
\label{fig:eval-megamoe}
\end{figure*}

E1 replays the mixed trace of Section~\ref{sec:bg-imbalance} at 8,192 tokens/rank through the placement, quota repair and local-first route rewrite of \mbox{HALO-Q}, planning each (iteration, layer) independently; it counts assignments, not time.
A CPU reference of the \mbox{HALO-Q} kernel and an earlier GB200 build of it matched the simulator bit for bit; the kernel at commit \texttt{99dac8f1} has not been checked.

\textbf{Objective 1, balance.}
With $S = 4$, \mbox{HALO-Q} lowered the 61-layer mean of $R$ from 1.20/1.44/1.84/2.24 to 1.002/1.006/1.010/1.020 at EP4/8/16/24, removing 98.4--99.0\% of the excess load ($S = 3$: 94.9--95.9\%), and kept every layer within 13\% of balance (Figure~\ref{fig:eval-placement}(a)).
An earlier policy that replicates four experts on every rank, which also produced the routes of E3, left 1.038/1.193/1.348 at EP8/16/24 with 28/60/92 remote copies per iteration and layer, against \mbox{HALO-Q}'s 20.4/41.4/76.2.

\textbf{Objective 2, locality.}
The most routed tokens that any rank sent out fell by 2--8\%, and the most that any rank received in by 19--55\% (Figure~\ref{fig:eval-placement}(b)).
Balance does most of the latter, as hot owners no longer receive the excess; quota repair alone cut both maxima by a further 0.5--2.3\%.

\textbf{Static plan.}
A static per-layer plan with the same four extra slots per rank, built by our re-implementation of the EPLB rule~\cite{deepseek-eplb} from past or whole-trace loads under an optimistic even token split, left $R = 1.06$--$1.07$ at EP8, against 1.006 with per-invocation planning.
By the GB200 fit of Section~\ref{sec:bg-imbalance} ($513+1051\,R$\us), extrapolated below its fitted range and without helper-tile overhead, a static plan would capture 85--88\% of the kernel-time gain of per-invocation planning over no balancing.

\textbf{Alignment and \sami.}
A broadcast charges a helper slot on every group member; the chosen group excluded the owner in 53\% of broadcasts at EP8 and 76\% at EP16--24.
Only \sami serves such groups (Section~\ref{sec:bg-ep}); restricted to groups that contain the owner, $R$ rose to 1.038/1.201/1.296 at EP8/16/24.

\textbf{Per-rank point-to-point placement.}
The point-to-point planner of Motivation~2 (Section~\ref{sec:mot-owner}) avoids that charge: with at most four copies per owner and four replicas per rank, it made 2.5--3.3$\times$ fewer remote copies than \mbox{HALO-Q} (EP8--24) but balanced worse.

\subsection{E2: \sami Broadcast (Reference Data)}\label{sec:eval-bcast}

E2 compares \sami with the single-port and ring broadcasts of Section~\ref{sec:mot-structures}, using a CE prototype, which issues the steps of Section~\ref{sec:design-sami} as host CE copies of one contiguous share per member, and two other builds of the replication kernel in standalone harnesses.
One owner broadcasts bundles to $g$ GPUs, itself included except in (c)'s TMA build.

\textbf{Against single-port multicast.}
In our placements, a single injecting port lost per-receiver bandwidth as $g$ grew, whereas \sami stayed near the owner's egress limit (Figure~\ref{fig:eval-bcast}(a)): its speedup grew from 1.13$\times$ at $g = 4$ to 2.03$\times$ at $g = 8$ and 3.76$\times$ at $g = 16$.
At $g = 4$, \mbox{HALO-Q}'s group size at EP8 and EP16 in replay, one port's 535\GBps per receiver is already close to that limit.

\textbf{SM-issued copy.}
The implementation's TMA copies add a per-SM issue limit.
Figures~\ref{fig:eval-bcast}(b, d) use a development build of the replication kernel (four SMs by default).
Standalone with eight bundles (283.5\,MiB; Figure~\ref{fig:eval-bcast}(b)), owner-direct multicast was issue-bound, rising from 46\GBps on one SM to 258\GBps on eight, whereas \sami, with every member issuing a share, reached 481\GBps from four SMs on (with four bundles, 166 and 450\GBps on four SMs).
The default per-slot rule (Section~\ref{sec:design-sami}) nonetheless picks owner-direct here: 258\GBps on eight SMs.

\textbf{Against ncclBroadcast.}
Figure~\ref{fig:eval-bcast}(c) completes each broadcast on every rank before the next starts and compares the fastest tuned ncclBroadcast setting of \mbox{nccl-tests} with the fastest \sami variant, mostly another TMA build of the replication kernel with the owner outside the multicast group (Section~\ref{sec:design-sami}).
We tuned NCCL~2.31.2 at every point over the protocol, the buffer size, the channel count, buffer registration and CUDA-graph launch; the LL128 protocol with CUDA-graph launch won all 12 points and cut the default ring's time (Simple protocol, 512-KiB slices) by 12--70\%.
\sami was 1.07/1.21/1.34$\times$ faster for one expert (35.4\,MiB) at 8/12/16 GPUs and 1.22--1.29$\times$ for eight: it took 101--107\us per expert at every $g$, against 108--143\us for the tuned ring, which still grows with $g$ (the default ring by 20\us per added GPU).
For eight experts on 8 GPUs, this variant reached 591\GBps against 486\GBps in (b), where the owner, a group member, also receives every replica, which slows the pulls it serves; (b)'s \sami matched tuned ncclBroadcast there (1.00$\times$).

\textbf{Overlap with the shared expert.}
Figure~\ref{fig:eval-bcast}(d) runs the four-expert copy (EP8; balanced synthetic routing, dummy weights) during a $7168{\rightarrow}2048{\rightarrow}7168$ FP8 shared expert, narrower than the model's.
With \sami, the copy took 310--330\us and lengthened the span by 2.0/0.6/0.3\% at 8K/16K/32K tokens/rank; with owner-direct multicast, it took 0.98--1.20\,ms and lengthened the span by 36.3/14.7/2.9\%, mostly where the copy outlasted the shared expert (256/480/936\us; Section~\ref{sec:mot-budget}).
Because every rank copies a share, \sami slowed the shared expert by 10/4.8/2.0\% and \megamoe by at most 0.5\%.
With the replication kernel of \texttt{99dac8f1} on eight SMs, \sami slowed the 8K shared expert by 49\%: in the profile, the copy ran beside the shared expert's GEMMs for 99.9\% of its time rather than in launch gaps, and the GEMMs, with no SMs set aside for the copy, slowed instead; the SM-free CE prototype slowed it by 2.9/1.8/0.9\%.
End to end on pre-release Rubin GPUs (EP8), a fused shared expert that leaves the copy's four SMs free ran 156--161\us per layer beside the copy, against 153--158\us without copies, and the copy spent 71--80\% of its time beside it and 2.5--5.6\% in idle gaps (medians per rank). %

\subsection{E3: \megamoe Kernel Time}\label{sec:eval-megamoe}\label{sec:eval-kernel}

We time one call of a later \megamoe revision than \texttt{99dac8f1}'s for a DeepSeek-V4-Pro layer on GB200, OFF and ON with four helper slots per rank, under two routings: synthetic power-law routing at four shapes (EP4 and EP8; 8,192 or 16,384 tokens/rank), each with a balanced control, and per-layer routes from the trace.
All runs used FP4 combine instead of the default BF16 combine.
ON's replicas, routes (from E1's all-rank policy) and READY flags were installed before timing, so the times exclude \mbox{HALO-Q}, the copy and any READY wait.
Each cell or layer ran as OFF, ON, OFF (OFF: the mean of its two runs); outputs matched bitwise.

\textbf{Power-law routing.}
With power-law popularity (exponent 0.8), the replicas reduced the skewed cells' $R$ from 1.27--1.69 to 1.00--1.03 and kernel time by 17.2--34.1\% (1.21--1.52$\times$), to within 0.3--3.5\% of balanced routing (Figure~\ref{fig:eval-megamoe}(a)); on balanced routing, time changed by $-0.7$\% to $+0.4$\%.\footnote{We report medians because single outlier iterations can dominate the mean of 20, the runs' declared statistic; with the mean, the skewed cells took 17.4--34.0\% less time and the balanced cells changed by $-5.0$\% to $-0.3$\%, the extremes due to single slow OFF iterations.}

\textbf{Trace routing.}
For each of the 61 MoE layers, we sampled EP8 routes at 8,192 tokens/rank from the layer's expert popularity in the mixed trace, calibrated to its EP8 imbalance within 3\% (one seed).
Every layer improved, from 1.3\% (layer 45, $R = 1.12$) to 42.8\% (layer 20, $R = 2.36$), and the mean time fell from 1,999.3 to 1,624.3\us (1.23$\times$), 2.1\% above balanced routing (Figure~\ref{fig:eval-megamoe}(b)): OFF time grew linearly with $R$ (Section~\ref{sec:bg-imbalance}), whereas ON time was uncorrelated with it (Pearson 0.01).
Both routings are stationary, so E3 measures the kernel's gain from replicas, not the value of per-invocation planning (Section~\ref{sec:eval-placement}).
ON's slowest layers carry the largest imbalance left by all-rank replication, which \mbox{HALO-Q} reduces in replay only (worst EP8 layer 1.066 vs.\ 1.198).

\textbf{Costs outside the kernel.}
In a sequential GB200 run of an earlier build, with ranks released together, \mbox{HALO-Q} took 19.6--29.2\us and the replication kernel (8 SMs, including READY) 128.7--340.6\us; serialized, the copy would cost 31--63\% of the kernel time saved on skewed routing (derived across runs; indicative), so it must overlap other work.
Planning time grows with the EP degree: on one CTA, placement and quota repair took 3.8, 6.1, 8.7 and 12.2\us at EP4, 8, 16 and 32 (medians on the trace's routing counts, three helper slots, four at EP4; pre-release Rubin GPU), as the greedy placement makes about 19 sequential exports per invocation at EP32 against 5 at EP8. %
In an overlapped run of an earlier build beside a 300\us single-CTA delay kernel standing in for the shared expert, with ranks arriving 10\us apart, ON took 15.6--29.2\% less time than OFF on the skewed cells and 0.2--1.5\% more on the balanced ones, planning, copy and READY waits included.
The stand-in moves no data, so this run misses interference with the real shared expert.

\subsection{E4: End-to-End Prefill Serving}\label{sec:eval-e2e}

\begin{figure}[t]
\centering
\includegraphics[width=\columnwidth]{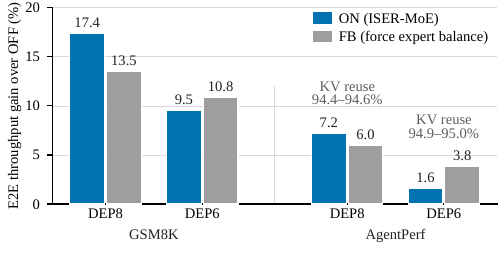}
\caption{E4 on GB300: end-to-end context-only throughput gain over OFF of ON (\mbox{\sys}) and of FB (force expert balance: OFF with routing forced to be perfectly balanced); DeepSeek-V4-Pro NVFP4, tile-wise FP4 combine. AgentPerf labels: key-value (KV) cache reuse rate across the three arms.}
\label{fig:eval-gb300}
\end{figure}

We served DeepSeek-V4-Pro (NVFP4) context-only in TensorRT-LLM with attention DP at EP8 or EP6 (DEP8, DEP6), OFF and with \sys (ON); throughput counts computed (non-reused) tokens per GPU and second.
On GB300, a third arm, force expert balance (FB), ran OFF with TensorRT-LLM's perfect router, which forces synthetic, perfectly balanced routing; its outputs are not meaningful.
All GB300 arms used tile-wise FP4 combine, in which the second GEMM's epilogue quantizes each output tile to NVFP4.
On GSM8K~\cite{gsm8k}, ON raised end-to-end throughput by 17.4\% at DEP8 and 9.5\% at DEP6, against FB's 13.5\% and 10.8\% (Figure~\ref{fig:eval-gb300}).
On AgentPerf~\cite{aa-agentperf}, with 94.4--95.0\% key-value (KV) cache reuse in every arm, ON gained 7.2\% and 1.6\%, against FB's 6.0\% and 3.8\%.
These gains come from load balance; multicast keeps the busiest owner at four or fewer send-outs per layer, where point-to-point copies under the same greedy placement need up to 10--14 at EP8--24 (Section~\ref{sec:mot-owner}); the implementation has no unicast copy path, so we did not compare the two end to end.

\begin{figure}[t] %
\centering %
\includegraphics[width=\columnwidth]{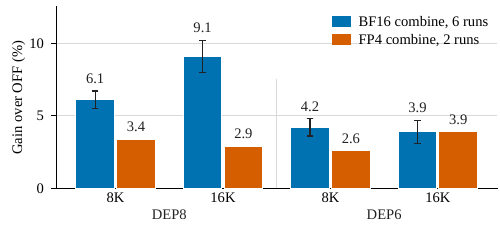} %
\caption{Context-only benchmark on pre-release Rubin GPUs (DeepSeek-V4-Pro NVFP4, attention DP; 8K and 16K: at most 8,192 and 16,384 tokens/rank per iteration): %
gain of ON over OFF in computed-token throughput on AgentPerf prompts (whiskers: the reported $\pm$). Every configuration gained.} %
\label{fig:eval-e2e} %
\end{figure} %

On pre-release Rubin GPUs, OFF and ON ran with at most 8,192 or 16,384 tokens/rank per iteration. %
On AgentPerf requests (95.1--95.9\% KV cache reuse), throughput was measured in six runs per arm with BF16 combine, the implementation's default, and in two with FP4 combine. %
With BF16 combine, throughput rose 6.1\% and 9.1\% at EP8 (8,192 and 16,384 tokens/rank) and 4.2\% and 3.9\% at EP6 ($\pm$0.6--1.1 points; Figure~\ref{fig:eval-e2e}), and time to first token (TTFT) improved 6.6--12.1\% except at EP6 with 8,192 tokens/rank ($-1.2\pm2.4$\%); with FP4 combine, throughput rose 2.6--3.9\%. %
A profile of an earlier commit (one rank, EP8, BF16 combine, $S=3$) shows the routed \megamoe kernel 29.5\% and all MoE work 15.0\% shorter per layer. %
The shared expert, beside the copy and with SMs held back for it, slowed by 127\us (from 182.6\us); the profile does not separate the two causes. %

\subsection{E5: End-to-End Disaggregated Serving}\label{sec:eval-disagg}

\begin{figure}[t]
\centering
\includegraphics[width=\columnwidth]{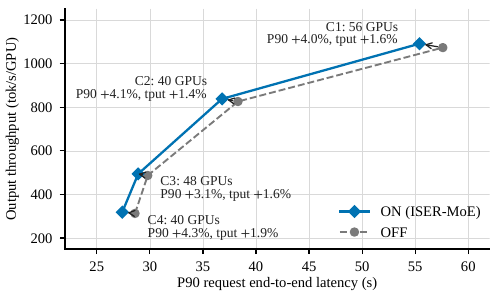}
\caption{E5 on GB300 NVL72: disaggregated serving of DeepSeek-V4-Pro (NVFP4) on agentic coding traces, output throughput per GPU against P90 request end-to-end latency, OFF and ON (\mbox{\sys}), configurations C1--C4 joined in order of concurrency. Labels: improvement of ON, OFF/ON$-$1 for latency and ON/OFF$-$1 for throughput (tput); each arm's best run.}
\label{fig:eval-disagg}
\end{figure}

We served DeepSeek-V4-Pro (NVFP4) in TensorRT-LLM with prefill and decode disaggregated on a GB300 NVL72 rack, in four configurations: C1 to C4 ran five, three, two and one DEP8 prefill instances with one DEP16 (C1, C2) or DEP32 (C3, C4) decode instance, on 56, 40, 48 and 40 GPUs at 2,626, 1,152, 736 and 388 concurrent requests.
Decoding used speculative decoding with three draft tokens and acceptance fixed at 2.01.
Clients replayed agentic coding traces~\cite{sa-cc-traces} with aiperf for a 1,200\,s window after warm-up; ON used three helper slots per rank, and KV cache reuse was 94.6--95.2\%, equal in both arms of every configuration.
In C3 and C4, ON also kept replicas resident: a replication whose whole group already holds the expert in that helper slot moves no data; C4 also skipped replications that would move fewer than 256 routes.
C1--C3 split the prefill instances between the arms, which shared the decode instance, in each of four, four and two jobs on the same nodes, alternating positions; C4 ran each arm in its own job, as two ON/OFF pairs on different nodes. Because serving performance varies between runs, each arm repeated one fixed configuration and reports its best run.
ON shortened P90 request latency by 4.0, 4.1, 3.1 and 4.3\% (OFF/ON$-$1) and raised output throughput per GPU by 1.6, 1.4, 1.6 and 1.9\% in C1--C4 (Figure~\ref{fig:eval-disagg}), so every configuration moved toward lower latency and higher throughput.

\subsection{Limitations}\label{sec:limits}

On GB200 we have not measured a layer with the real shared expert or end-to-end throughput, and no hardware result compares with a static plan (Section~\ref{sec:eval-placement}).
Each member of an aligned group spends a helper slot and receives the whole expert, and owner-direct replication was issue-bound on 8 SMs in a development build of the replication kernel (E2).
Without a shared expert, only the home tiles can hide the copy, which leaves room for at most about one DeepSeek-V4-Pro-sized copy (GB300 peak-rate model, EP8, 8,192 tokens/rank) and none with a BF16 combine (Section~\ref{sec:mot-budget}).
The scope is prefill with attention DP in one NVLink domain, in eager mode; the implementation accepts up to EP32 (Section~\ref{sec:impl}), but hardware runs cover EP4--8 and replay EP4--24, and no controlled hardware point lies below 8,192 tokens/rank.
At EP32, planning takes 12\us, twice its time at EP8 (Section~\ref{sec:eval-megamoe}). %
Replicas live for one invocation, although in replay 74--78\% of the all-rank policy's expert selections recur in the next iteration; placement cannot fix ranks that reach the layer at different times.

\section{Related Work}\label{sec:related}

\textbf{Static replication.}
Deployed systems replicate hot experts from load history: DeepSeek-V3 periodically adjusts redundant prefill experts~\cite{deepseekv3}, EPLB plans replicas from estimated loads~\cite{deepseek-eplb}, and serving engines move weights between rebalances~\cite{trtllm-blog-wideep,vllm-megamoe-eplb}.
Such plans can worsen imbalance as load shifts~\cite{ultraep}; the Linear-Programming-Based Load Balancer (LPLB)~\cite{deepseek-lplb} re-splits each batch's tokens over static replicas without moving weights, so it corrects drift only for replicated experts.
METRO~\cite{metro} routes over static replicas to minimize activated experts in memory-bound decode, not compute-bound prefill.
Static plans and per-invocation planning are complementary (Section~\ref{sec:eval-placement}).

\textbf{Per-invocation replication.}
MegaFlux~\cite{megaflux}, concurrent with our work, replicates hot experts at every invocation inside the \megamoe kernel that we extend and pipelines the transfer with per-projection readiness, hiding an estimated 56--76\% of it; it describes neither multicast nor relays, so by our reading an owner sends one transfer per replica.
UltraEP~\cite{ultraep} rebalances every layer and micro-batch from the exact load, identifies hot-expert fan-out as the weight-distribution bottleneck, relays experts with many replicas in two stages, and sets in-switch multicast aside because replica groups change; we keep that premise and fix the groups instead, as FasterMoE broadcasts to one fixed group~\cite{fastermoe}.
MoonEP~\cite{moonep}, described in the Kimi~K3 report~\cite{kimik3}, balances training load exactly with $E/P$ prefetch slots per rank, each owner pushing weights with one TMA store per destination; ECHO (Elastic Cloning for Hot Experts) and the concurrent DeepEP~V2.5 communication library also replicate experts dynamically~\cite{megatron-moe,deepep}.
All of them copy replica weights point to point; \sys multicasts them and accepts a small residual imbalance.

\textbf{Hiding the copy.}
FEPLB and TAOT (Topology-Aware Optimal Transport) hide replica copies behind home-expert computation~\cite{feplb,taot}, and PROBE behind earlier computation by predicting routing~\cite{probe}; Megatron-Core and Kimi~K3 overlap the shared expert with token communication~\cite{megatron-moe,kimik3}, but, to our knowledge, no prior system that replicates experts per invocation uses it to hide planning and replication.

\textbf{Reported gains.}
Reported gains are not directly comparable: MegaFlux reports 1.45$\times$ at kernel level, including planning and transfers that our 1.21--1.52$\times$ excludes, and 1.13--1.26$\times$ end to end in vLLM prefill of DeepSeek-V4-Pro; UltraEP reports 1.29$\times$ over SGLang with EPLB at EP40--64.

\textbf{In-network collectives.}
NVLS accelerates collectives over fixed groups~\cite{nccl,nvswitch-hotchips22} and its AllGather and two-shot all-reduce already inject from every member's port~\cite{nccl,trtllm-multishot}; \sami applies this to weight replication, with an owner that may lie outside the group.
MoE systems multicast plan metadata~\cite{moonep}, tokens~\cite{megatron-nvls-dispatcher}, activations~\cite{kimik3} and bucket collectives~\cite{deepep} in the switch, and DySHARP proposes switch support for token multicast~\cite{dysharp}; none multicasts expert weights.

\section{Conclusion}\label{sec:conclusion}

Per-invocation replicas of hot experts can remove most of the rank imbalance in expert-parallel MoE layers, but only if their weights arrive within the layer.
\sys multicasts them through the NVLink switch to aligned GPU groups created once, so each byte leaves its owner once; \sami spreads the injection over the group's ports, and planning and replication overlap the shared expert.
In trace replay, \mbox{HALO-Q} removed 98--99\% of the excess load, and on GB300 at EP8, context-only throughput rose 17.4\% on GSM8K and 7.2\% on agentic prompts.
On pre-release Rubin GPUs at EP8, context-only throughput rose 6.1\% on agentic prompts. %
A hardware comparison with static plans, generation and training, attention-side imbalance and replication across NVLink domains remain future work.

{\footnotesize
\let\url\nolinkurl
\setlength{\bibsep}{0pt plus 0.5pt}
\bibliographystyle{abbrvnat}
\bibliography{refs}
}

\end{document}